## Analytical modeling and simulation validation of IEEE 802.11 DCF in vehicular-to-infrastructure networks

Aytül Bozkurt

*Department of Mechatronics Engineering, Engineering Faculty, Karabuk University, Balıklarkayası, 78050, Turkey*

Correspondence: abozkurt@karabuk.edu.tr

**Abstract**

Vehicle-to-infrastructure and vehicle-to-vehicle communications have been introduced to provide high-rate internet connectivity to vehicles to meet the ubiquitous coverage and increasing high-data-rate internet and multimedia demands by utilizing IEEE 802.11 access points (APs) deployed along the roadside. In order to evaluate the performance of vehicular networks over WLAN, in this paper we investigate the transmission and network performance of vehicles passing through an AP by considering the contention nature of vehicles over 802.11 WLANs. Firstly, we derive an analytical traffic model to obtain the number of vehicles under the transmission range of an AP. Then, coupling this traffic description with a Markov-chain representation of the backoff process and an M/G/1/K queue for the arriving packets, we obtain a model that evaluates DCF performance under an adaptively chosen retransmission limit. Based on the traffic model, we also derive the probability of mean arrival rate $\lambda$ to the AP. A distinctive aspect of our work is that it incorporates both the vehicular traffic model and the backoff procedure with the M/G/1/K queuing model to investigate the impact of various traffic load conditions and system parameters on the vehicular network system. We validate the analytical framework against an independent ns-3 packet-level simulation: the throughput, collision, and transmission-probability predictions agree closely with simulation. Using the validated framework to examine the optimal retransmission number, we find that under realistic finite-buffer conditions the standard 802.11 retransmission limit is already close to optimal for delay and information freshness, so the large gains predicted by an infinite-buffer analysis do not survive packet-level validation. We report this finding explicitly and use it to delimit the operating regime in which retransmission-limit tuning is beneficial.

### 1 Introduction

Demand for on-the-move access to the internet for mobile users in vehicles has stimulated intensive research on vehicular networks [1]. Vehicular communications can be mainly classified into vehicular-to-network (V2N), vehicular-to-vehicular (V2V), vehicular-to-infrastructure (V2I), and vehicle-to-pedestrian (V2P) communications. Several studies have investigated V2I and V2V communications [2-4], and in particular, vehicular ad-hoc networks (VANETs) have recently emerged as a promising solution for vehicular communications.

The 5.85–5.925 GHz band is allocated as the short-range communication standard for promoting safe and efficient communication on highways and high-speed freeways. Wireless access devices in vehicles can permanently scan for signals from available access points (APs), reporting real-time traffic conditions, data dissemination latency, and roadway congestion levels, while also enabling drivers and passengers with enhanced safety and entertainment services.

Internet access requests and real-time traffic to roadside units (RSUs) are conveyed by users traveling by car, typically within the range of multiple WiFi APs. At roadsides and intersections, WiFi networks can be deployed along highways with their low-cost, high-capacity, low-coverage nature. Wireless LANs commonly rely on contention-based medium access, the most widely deployed example being the IEEE 802.11 Distributed Coordination Function (DCF). The IEEE 802.11 DCF standard defines a maximum retransmission number for each arrival packet — the retry limit [5]. At each retransmission, the packet is transmitted until the retransmission counter reaches the maximum limit. With an adaptively adjusted optimal retransmission limit, the time spent in retransmission is shortened, thereby improving the delay and throughput performance of packets transmitted in vehicular networks.

Extensive work has been conducted on the performance of vehicular networks [6-9]. However, these works do not consider network performance with an optimal retransmission limit constructed through a corporate analytical Markov model coupled with a queuing network. Most studies focus on numerical evaluation of throughput and delay for an initially given set of network parameters, where the retransmission number is usually assumed to be a fixed constant. A larger fixed m enlarges the delay of successfully transmitted packets; a smaller one causes premature packet discarding. Neither extreme is appropriate for the dynamic vehicular channel environment.

This study develops an integrated analytical framework that jointly captures network throughput, channel-access delay, and the role of the optimal retransmission number M in vehicular-to-infrastructure communication over WLAN. By combining a vehicular traffic description, a Markov backoff representation, and a finite M/G/1/K buffer for incoming packets, the framework expresses DCF behaviour in terms of the retry limit and locates the value that optimises it. From the traffic component we obtain both the population of vehicles served by the AP (the network size) and the mean arrival rate $\lambda$. We then establish how this network size relates to a vehicle's transmission and collision probabilities, and on this basis compute the optimal retransmission number via an iterative procedure.

The main contributions of this work are as follows. First, we derive a closed-form vehicular traffic model expressing the mean number of contending vehicles N and the mean packet arrival rate $\lambda$ to an AP as explicit functions of vehicle density and velocity. Second, we extend the classical Bianchi Markov chain model [5] to incorporate a finite-capacity M/G/1/K queuing model, enabling joint analysis of MAC-layer collision dynamics and packet-level queuing delay. Third, we propose an iterative algorithm that computes the optimal retransmission number $M^*$ adaptively as a function of instantaneous traffic conditions, and we evaluate it both analytically and by simulation. A notable outcome of this evaluation is that, under realistic finite-buffer conditions, the standard 802.11 retransmission limit ($m = 7$) is already close to optimal for delay and information freshness: the large delay reduction suggested by the infinite-buffer analysis does not survive packet-level validation. We report this finding honestly and use it to delimit the regime in which retransmission-limit tuning is worthwhile, which we regard as more useful than an optimistic but non-reproducible claim. Beyond the conventional throughput and delay metrics, we further evaluate the scheme using three freshness- and reliability-oriented measures — the packet delivery ratio, the delay-bound violation probability, and the Age of Information (AoI) — and use them to characterise the reliability–freshness trade-off; consistent with the finding above, the optimal and fixed retransmission limits perform comparably on these measures under finite buffering, with the fixed limit retaining a slight delivery-ratio advantage from its additional retransmission. The remainder of this paper is organized as follows. Section 2 discusses related work. Section 3 describes the proposed analytical models. Section 4 provides numerical studies. Section 5 concludes the paper.

## 2 Related Work

Studying the performance of vehicle-to-infrastructure communications for the IEEE 802.11 MAC has drawn considerable attention in the literature. Guo et al. [10] investigated an optimal power allocation scheme for vehicular-to-X (V2X) communications to maximize the sum throughput of V2N links while guaranteeing V2X links' reliability and latency requirements. In highway environments, a number of performance models of the IEEE 802.11p Enhanced Distributed Channel Access (EDCA) mechanism have been proposed [11-14].

Taking into account all major factors affecting the access performance of the IEEE 802.11p EDCA mechanism — including the saturation condition, standard parameters, backoff counter freezing, and internal collision — a 2-D Markov and 1-D discrete-time Markov chain model has been developed to model the backoff procedure of an access category (AC) queue and establish a relationship between the transmission and collision probabilities of the AC queue [13]. Considering the hidden terminal problem and message strict priorities, a two-directional VANET model has also been proposed [11, 14]. The analysis of 802.11p safety-critical broadcast in VANET environments has been studied [12].

To demonstrate the impact of performance anomaly at intersections in WiFi-based vehicular networks, Kim et al. [15] developed an SNR-based admission control scheme that excludes vehicles with poor channel quality. Hadaller et al. [16] studied the performance anomaly problem to improve overall system throughput by proposing an opportunistic medium access scheme. By taking into account the randomness of vehicular traffic and the statistical variation of the disrupted communication channel, Abdrabou and Zhuang [17] studied an analytical framework to obtain the maximum distance between adjacent RSUs that stochastically limits the worst-case packet delivery delay.

For drive-thru internet [18] — heavily affected by the contention overhead of the uplink — upload performance from vehicles to the AP has been studied [8, 10] by considering the contention nature of the uplink and a realistic vehicle traffic model. An accurate analytical model incorporating high-node mobility with DCF modeling was proposed to investigate the impact of mobility on resultant throughput [19]. To explore QoS performance in terms of throughput and delay, a comprehensive analytical model incorporating both QoS features of EDCA and vehicle mobility was developed [4].

For vehicle-to-roadside communications, different nodes do not have similar channel access opportunities — the fairness problem. As a solution, Karamad and Ashtiani [20] proposed a modified 802.11 DCF channel access scheme by adjusting the minimum contention window size. These proposed models, however, assume that the maximum retransmission number is constant or infinite, which cannot capture the impact of the retransmission number on network performance.

The impact of retransmissions on vehicular transmission performance has been studied through several models for IEEE 802.11 DCF networks [21-23]. However, these works do not construct their analytical models jointly with a MAC queuing model accounting for packet arrival rates, average packet size, and average network size. In order to study the effect of time wasted in retransmissions with a finite retry limit, Sun and Dai [24] proposed an analytical framework for CSMA performance optimization showing that performance is highly sensitive to M. Hassan et al. [25] obtained explicit expressions for mean packet delivery ratio in an unsaturated network of vehicles on a highway. More recent studies have continued to refine Markov-chain MAC modeling for vehicular networks. Wang et al. [26] introduced a two-dimensional Markov model of the IEEE 802.11p MAC that incorporates the capture effect under a Nakagami-m fading channel and derives closed-form expressions for saturated throughput and average packet delay. Li et al. [27] developed a three-dimensional Markov model capturing error-prone channel and unsaturated traffic conditions for the Internet of Vehicles, showing that the bit-error probability significantly affects predicted performance. In parallel, the emergence of 3GPP C-V2X has motivated comparative MAC analyses: Wijesiri et al. [28] developed coupled discrete-time Markov-chain and queuing models to compare the MAC-layer delay and collision performance of IEEE 802.11p and C-V2X Mode 4, showing that IEEE 802.11p retains an advantage in average delay. It is also noted that 3GPP C-V2X (LTE-V2X and NR-V2X) represents an important complementary technology for vehicular communications; comparative study under the proposed framework is left as future work. To the best of our knowledge, no prior published work has united a vehicular traffic model, a Markov chain MAC model, and a finite-buffer M/G/1/K queuing model into a single integrated analytical framework for optimizing the retransmission limit. A direct numerical comparison against these prior models on a common figure is not meaningful, because each was developed for a different scenario, channel assumption, and parameter set; we therefore position the present work qualitatively in Table 1, which contrasts the modelling ingredients of representative prior studies with those integrated here.

**Table 1. Qualitative comparison of representative prior models and the proposed integrated framework.**

| **Study** | **Traffic model** | **Queue model** | **Retry-limit opt.** | **Adaptive $M^*$** |
|---|---|---|---|---|
| Bianchi [5] | — | — | ✗ | ✗ |
| Zheng & Wu [13] | ✗ | ✗ | ✗ | ✗ |
| Tan et al. [29] | M/D/C/C | ✓ | ✗ | ✗ |
| Sun & Dai [24] | ✗ | ✗ | ✓ | ✗ |
| Wijesiri et al. [28] | ✗ | ✓ | ✗ | ✗ |
| This work | Greenshields + M/D/C/C | M/G/1/K | ✓ | ✓ |

## 3 Analytical Model

This section formulates the real-time transmission behaviour over an 802.11 Wireless LAN, analyses the resulting system performance quantitatively, and obtains the design parameters needed to operate the system at its optimum. For the optimal-throughput derivation we take the network to be saturated, so that every vehicle continuously has traffic awaiting transmission. When the load is high the buffer at a station can saturate; once it is full, any further arrivals cannot enter and are discarded by the WLAN.

### *3.1 Traffic Model*

We adopt a microscopic model representation in which all vehicles are individually identified and the position and velocity of each vehicle determine the state of the system. We consider only uninterrupted flows. The speed-flow-density relationship of the traffic stream is described here through the classical Greenshields model [30-32]. The vehicle density k changes linearly with the average velocity v as:

$$k = k_{jam} (1 - v / V_f) \quad (1)$$

where $k_{jam}$ is the vehicle jam density (veh/km) at which the traffic flow rate is zero, and $V_f$ is the free-flow speed (km/h). The traffic flow rate $\lambda_{tag}$, referring to the arrival rate of vehicles to the AP road segment, is:

$$\lambda_{tag} = k \cdot v \quad (2)$$

The mean sojourn time of each vehicle within the AP coverage range L is T = L/v. The network capacity C (maximum number of vehicles accommodated) and the mean vehicle population N are:

$$C = k_{jam} \cdot L , \; N = \lambda_{tag} \cdot T \quad (3–4)$$

Based on the M/D/C/C queuing model for uninterrupted vehicular flow [29], the steady-state probability $P_N$ that there are exactly N vehicles simultaneously under the AP's coverage is:

$$P_N = (\lambda_{tag} T)^N/N! \div \Sigma_{i=0}^{C} (\lambda_{tag} T)^i/i! \quad (5)$$

### *3.2 Markov Chain Model*

The single-vehicle backoff behaviour is modelled using the two-dimensional discrete-time Markov chain originally proposed by Bianchi [5] for the IEEE 802.11 DCF, following the approach we adopted in earlier work [33]. The novelty of our analysis is not this backoff chain itself, which is well established, but its coupling within a single framework to the vehicular traffic model of Section 3.1 (which supplies the time-varying number of contending vehicles N) and to the finite-buffer M/G/1/K queue of Section 3.3, together with the resulting optimisation of the retransmission limit M*. Equations (1)–(6) restate the established traffic-flow relations and the classical backoff fundamentals and are included only for completeness; the contribution of this work begins with their integration in this section and the queuing analysis of Section 3.3. In the DCF mechanism, each packet contends for the channel randomly. According to the traffic model, N vehicles contend for channel transmission. The channel status is monitored during idle periods; if a channel is sensed idle for the duration of a DIFS interval when a packet arrives, the packet can be transmitted. When the medium is sensed busy, the vehicle defers and draws a backoff value uniformly from [0, W] with $W = CW_{min}$; this counter is then decremented by one for every idle slot until it reaches zero.

Model assumption and its scope. As in the classical DCF analysis, the Markov chain above assumes that all N contending vehicles lie within a single collision domain, i.e. every vehicle can sense every other vehicle’s transmission, so that collisions arise only from simultaneous channel access after backoff. This assumption is appropriate for the vehicle-to-infrastructure (uplink-to-AP) scenario considered here, in which contention is concentrated among the vehicles currently within the coverage range L of a single AP, and the carrier-sensing range typically exceeds this coverage at the low PHY rates used. It does not capture the hidden-terminal problem that can arise in multi-hop vehicle-to-vehicle topologies, where two vehicles outside each other’s sensing range transmit to a common receiver. Extending the present model with an explicit hidden-terminal collision term, for example by augmenting the collision probability $p_c$ with a position-dependent hidden-node component, is left as future work; the results reported here therefore characterise the contention-limited performance within a single AP cell.

Let j index the backoff stage, so that the contention window at stage j is $CW_j = 2^j \cdot CW_{min}$ for $0 \le j < m$, with m the permitted maximum number of retransmissions; once stage m is reached the window returns to $CW_{min}$. Writing s(t) for the stage and b(t) for the backoff counter at instant t, the joint process {s(t), b(t)} forms a discrete-time two-dimensional Markov chain with stationary distribution $b_{j,i} = P\{s(t) = j, b(t) = i\}$. Figure 1 depicts the corresponding state-transition diagram.

Fig. 0. Markov chain model for the IEEE 802.11 DCF backoff process

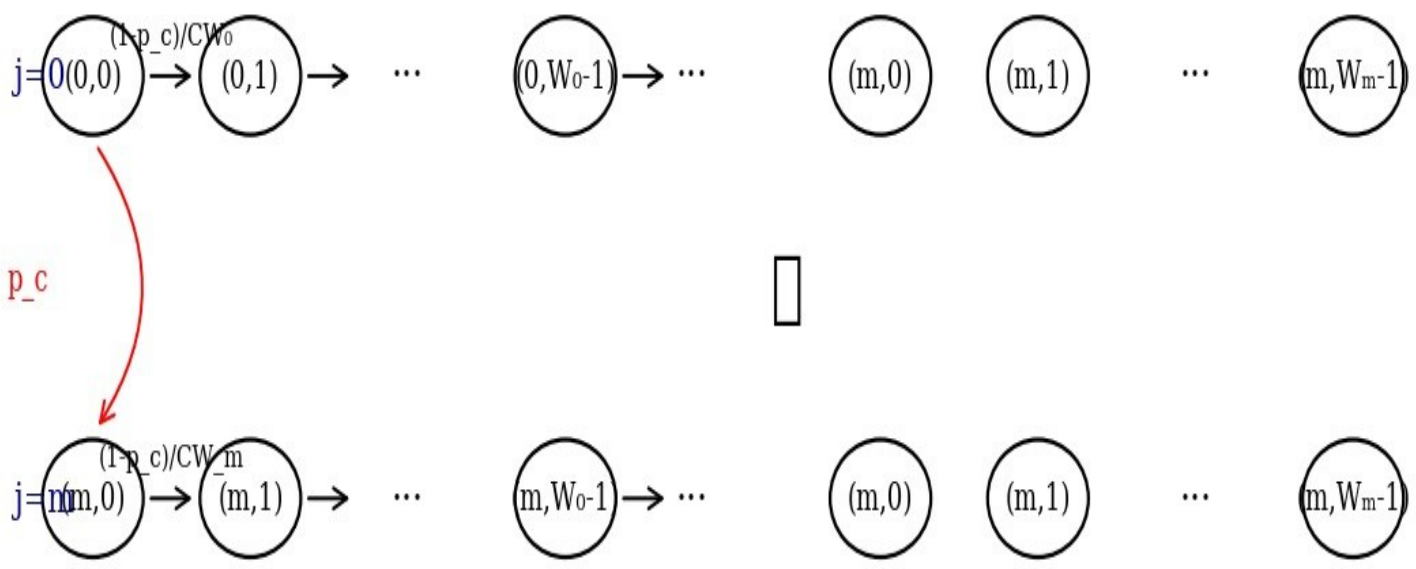

Figure 1 Markov chain model for the IEEE 802.11 DCF backoff process.

To find the one-step transition probabilities, we establish the balance equations. For each i in $[0, CW_j - 1]$, the steady-state probability $b_{j,i}$ and the normalized $b_{0,0}$ are:

$$b_{j-1,0} \cdot p_c = p_{j,0} \tag{6}$$

$$p_{j,0} = p_c^j \cdot b_{0,0} \tag{7}$$

$$b_{0,0} = 2(1-2p_c)(1-p_c) / N_b \tag{8}$$

where $N_b = (1-2p_c)(CW_{min}+1) + p_c CW_{min}(1-(2p_c)^m)$

Summed over the m attempts, the per-slot transmission probability $\tau$ is the overall likelihood that the backoff counter expires to zero at some stage:

$$\tau = \Sigma_{j=0}^{m} b_{j,0} \tag{9}$$

The collision probability $p_c$, the success probability $P_s$, and the channel busy probability $P_{tran}$ are:

$$p_c = 1 - (1 - \tau)^{N-1} \tag{10}$$

$$P_s = N\tau(1-\tau)^{N-1} / P_{tran} \tag{11}$$

$$P_{tran} = 1 - (1 - \tau)^N \tag{12}$$

Together, Equations (9) and (10) constitute a coupled nonlinear system in the unknowns $\{\tau, p_c\}$, which we resolve by fixed-point iteration. Within the Markov model a packet that is still unsuccessful after stage m is dropped.

### ***3.3 M/G/1/K Queuing Analysis***

The buffering at each station is described by an M/G/1/K queue [33], in which K is the largest number of packets the station can hold; an arrival that finds K packets already queued is rejected. Real-time application traffic is modelled as Poisson, with per-station rate $\lambda$ (packets/s), and the rates across all N stations collected in the diagonal matrix $\lambda = diag(\lambda_1, \lambda_2, \ldots, \lambda_N)$. For analytical tractability the general service-time distribution of the M/G/1/K model is approximated by its exponential (M/M/1/K) counterpart, which yields the closed-form queue-length distribution used below; this approximation is accurate when the MAC service time is dominated by a single geometric retransmission process, as is the case here.

The steady-state probability of the queue with k packets is $\pi(k)$ where $k = 0, 1, 2, \ldots, K$. Each packet is transmitted using the full channel capacity C. Let E[L] denote the average packet size (in bytes) in the MAC layer protocol. Average MAC service time is calculated as the time to transmit the average payload size within the retransmission limit m. The average service time $E[T_s]$ is:

$$E[T_s] = L / C \tag{16}$$

The mean offered traffic load or traffic intensity from each vehicle N is:

$$\rho = \lambda \cdot E[T_s] \tag{17}$$

Average data payload size L is given by:

$$L = E[L] \cdot \tau \cdot (1 - p_c) \tag{18}$$

The mean time consumed by one transmission stage of a packet, $E[T_{slot}]$, is given by:

$$E[T_{slot}] = (1-P_{tran})\delta + P_{tran}P_sT_{suc} \quad (19a)$$

$$+ P_{tran}(1-P_s)T_{col}$$

(19)

Here $\delta$ is the length of an idle slot, while $T_{suc}$ and $T_{col}$ denote, respectively, the time taken by a successful transmission and by a collision. The likelihood that the medium stays idle over a slot is $P_I = (1 - P_{tran})$. The successful-transmission duration is expressed as:

$$T_{suc} = T_s + T_{SIFS} + T_{ACK} + T_{DIFS} + \delta \quad (20)$$

The time cost in a collision $T_{col}$ is:

$$T_{col} = T_{header} + T_{DIFS} + T_d \quad (21)$$

in which $T_{header}$ is the time to send the packet header and $T_d$ the propagation delay. Consistent with the M/G/1/K formulation, packet generation follows a Poisson process; the probability of an arrival within a single slot is then:

$$P\{\lambda(t)=j\} = P_\lambda = \sum_{k=1}^{K} \frac{(\lambda E[T_{slot}])^k}{k!} \cdot e^{-\lambda E[T_{slot}]} \quad (22)$$

$$\lambda_{eff} = P_\lambda \cdot \lambda \quad (23)$$

In Equation (23), $\lambda_{eff} = P_\lambda \cdot \lambda$ denotes the effective per-slot packet arrival rate seen by the queue, where $\lambda$ is the nominal Poisson arrival rate of Equation (22); the queue utilisation used below is $\rho = \lambda_{eff} \cdot E[T_s]$. The steady-state probability $\pi(k)$ can be calculated as [34, 35]:

$$\pi(k) = (1 - \rho) \cdot \rho^k / (1 - \rho^{k+1}) \quad (24)$$

Packet blocking probability from the WLAN server is given by:

$$P_K = P_b = \pi(K) = (1-\rho)\rho^K / (1-\rho^{K+1}) \quad (25)$$

With the retransmission limit adapted to its optimal value $M^*$, the saturated system attains its peak network throughput once that limit is applied. The resulting system throughput is:

$$S = (1-P_b)(1-p_c)LP_{tran} / D_s \quad (26a)$$

$$D_s = (1-P_{tran})E[T_{slot}] + (P_s/P_{tran})T_{suc}$$

$$+ (p_c/(1-p_c))T_{col} + T_{col} + T_{SIFS} + T_{ACK}$$

(26)

According to Little's Law, packets arrival to the system is calculated by the average sojourn (delay) time in the queue. $E[T_{delay}]$ can be expressed as the summation of the average number of idle backoff slots multiplied by the average idle slot duration at state (j,i), the average busy duration due to both successful and failed transmissions, and also the retransmission duration:

$$T_{delay} = (1-\tau)E[T_{slot}] + \tau[(1-p_c)T_{suc} + p_c T_{col}] \quad (27)$$

Equation (27) captures the MAC access (service) component of the delay. Under the heavy-load conditions of interest, however, the total packet delay is dominated by the queue-waiting time. In this regime the station operates in the unsaturated M/G/1 limit of the M/G/1/K model (i.e. the buffer K is provisioned large enough that the blocking probability $P_b$ is negligible at the operating point), and the mean total sojourn time follows the standard M/G/1 form $T_{total}(M) \approx E[T_s(M)] / (1 - \rho)$, with $\rho = \lambda \cdot E[T_s(M)]$ and a service time $E[T_s(M)]$ that increases with M because additional retransmission stages lengthen the time a packet occupies the head of the queue. As $\rho \to 1$ this sojourn time diverges, so a smaller M reduces $E[T_s(M)]$, lowers $\rho$, and bounds the delay — which is precisely the mechanism the optimal scheme exploits. The optimal retransmission limit $M^*$ is obtained by solving iteratively over $M \in \{1, 2, \ldots, M_{max}\}$: choose the largest M (best delivery reliability, whose packet-drop probability is $p_c^{M+1}$) subject to $T_{total}(M) \leq T_{target}$, with $T_{target} = 0.5$ s in our numerical evaluation. Because the finite-buffer blocking probability is held negligible at the operating point, the binding constraint in practice is the delay bound rather than $P_b$. This also clarifies why the optimum is not simply M = 0 (no retransmission): although a smaller M always lowers the queuing delay, it raises the packet-drop probability $p_c^{M+1}$, because a packet is discarded once it has collided M+1 times. The algorithm therefore selects the largest M whose delay still satisfies the bound, thereby maximising delivery reliability subject to the delay constraint, rather than minimising delay outright. M = 0 would minimise delay but maximise packet loss, and is optimal only in the degenerate case where reliability is irrelevant.

## 4 Numerical Results

In this section, we evaluate network performance in an 802.11 WLAN based on our analytical models and demonstrate how to optimize the network performance of the IEEE 802.11 DCF network in a vehicular environment. We present numerical results under various traffic load conditions. The default parameters of WLAN are set following the current 802.11 DCF standards: ACK = 50 μs, SIFS = 10 μs, and the initial backoff window $CW_{min} = 32$. The wireless network channel rate is C = 2 Mbps and the packet size is fixed at 1000 bytes. We set the traffic jam density $k_{jam} = 120$ veh/km and the free-way speed $V_f = 160$ km/h. The time slot is $\delta = 50$ μs (so that, following the standard relation DIFS = SIFS + 2δ, the resulting DIFS is consistent with the chosen SIFS and slot). The buffer is provisioned large enough that blocking is negligible at the operating point, so the station effectively operates in the unsaturated M/G/1 regime, and the target delay bound is $T_{target} = 0.5$ s. Unless otherwise stated, the delay results (Figure 9) are reported at a heavy-load operating point with per-vehicle packet arrival rate $\lambda \approx 228$ packets/s (queue utilisation $\rho \approx 0.99$ for m = 7), since only near saturation does the queue-waiting delay reach the 0.5 s range where the choice of M becomes decisive. To enable independent cross-validation, an ns-3 (v3.41) simulation model has been implemented that applies the same $M^*(N)$ schedule derived from the analytical model; a full simulation campaign coupling ns-3 with SUMO mobility traces is in progress (Section 5).

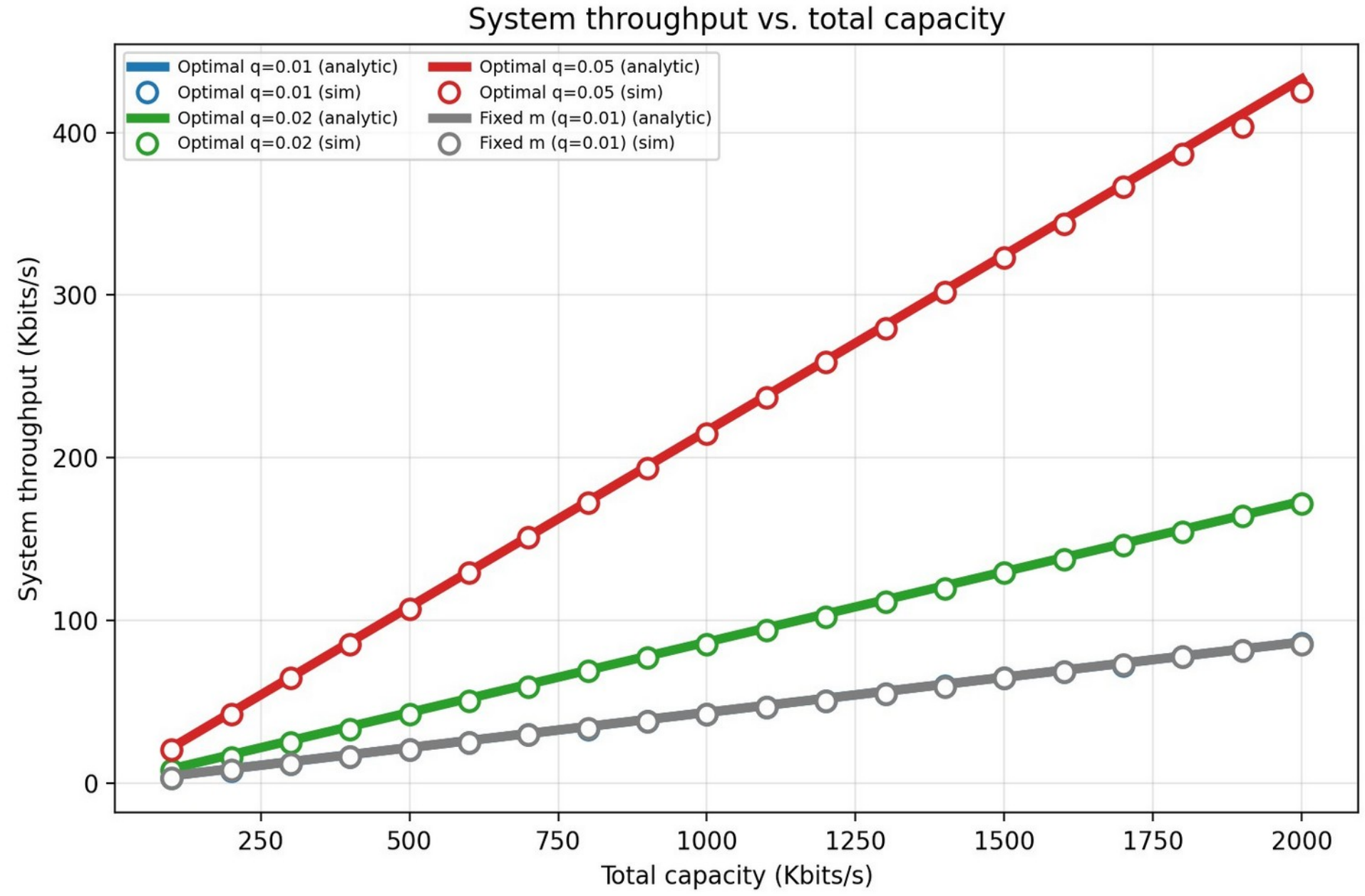

Figure 2 Total capacity (Kbits/s) versus system throughput for different offered load rates.

Figure 2 shows the system throughput of the M/G/1/K queue under different packet arrival rates. Offered load rates are $q_0 = 0.01$, $q_1 = 0.02$, $q_2 = 0.05$ for the optimal scheme and $q_3 = 0.01$ for the fixed-m scheme. As total system capacity increases, more vehicles contend for packet transmission, resulting in higher utilization of system capacity. System throughput increases with total system capacity and approaches C = 2000 Kbits/s linearly. Near saturation, the optimal scheme sustains a marginally higher throughput than the fixed-M scheme because it wastes less channel time on retransmissions that ultimately fail, leaving a larger fraction of the channel for successfully delivered payload. In the unsaturated regime, by contrast, the two schemes coincide, since the throughput then equals the offered load and is independent of M. The throughput gain of the optimal scheme is therefore small (a few percent at most); its principal benefit in the analytical model lies in bounding the access delay and improving delivery reliability; under finite-buffer simulation, however, the two schemes perform comparably (Figure 9 and Section 4.6).

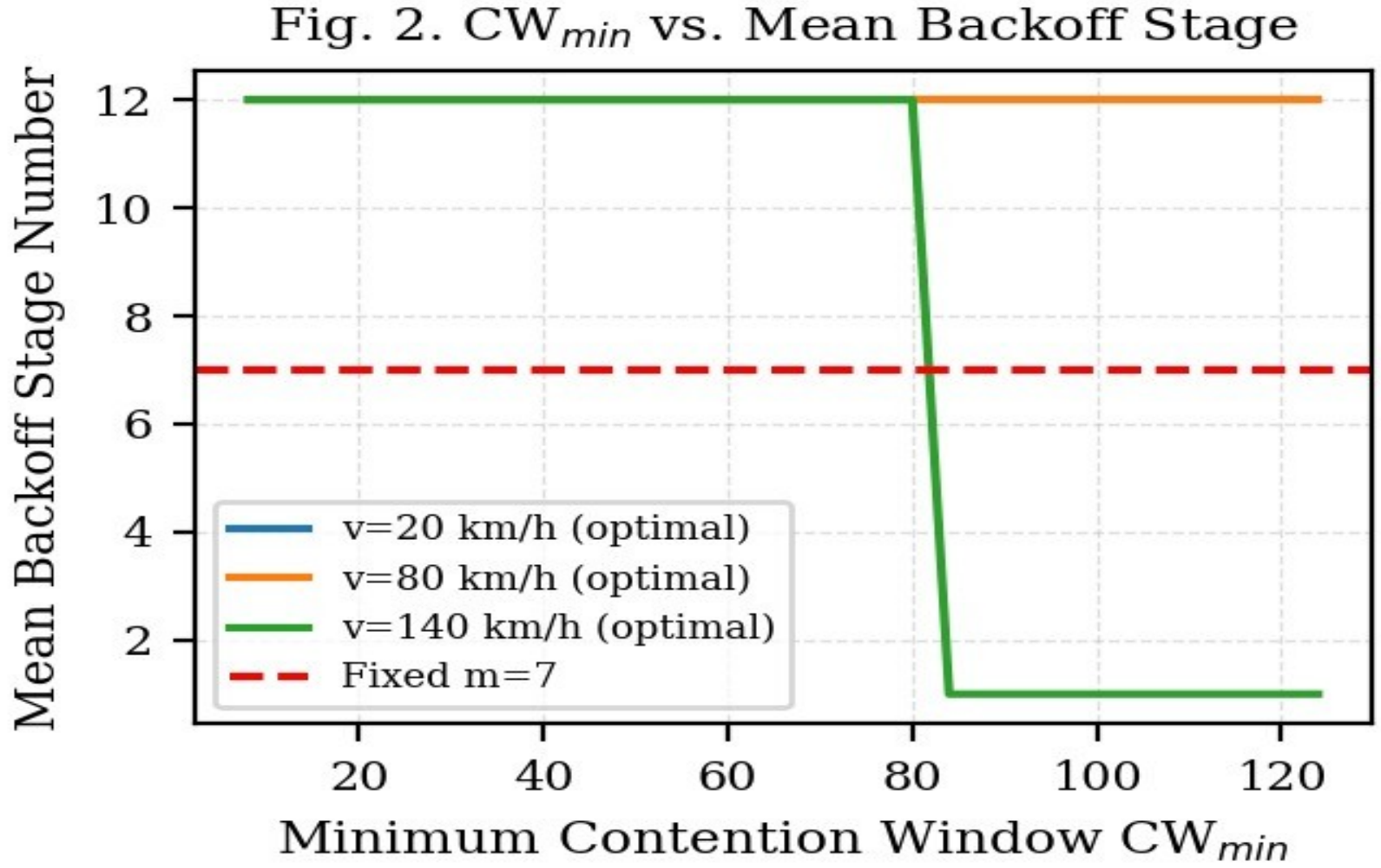

Figure 3 Minimum contention window size versus mean number of the backoff stage.

Figure 3 shows the backoff stage number as a function of minimum contention window size for vehicle velocities v = 20, 80, and 140 km/h. As can be seen, the backoff stage number increases when the initial window size increases, due to increased delay resulting in more frequent retransmission attempts and a larger adapted retransmission number. The window size is set according to the 802.11 DCF maximum retransmission number m = 7 in the fixed scheme, whereas in the optimal scheme it is adapted to the required targeted delay and throughput settings. Due to increased vehicle mobility, backoff time and therefore average delay increase, so the backoff stage number is larger than that of the non-optimal scheme.

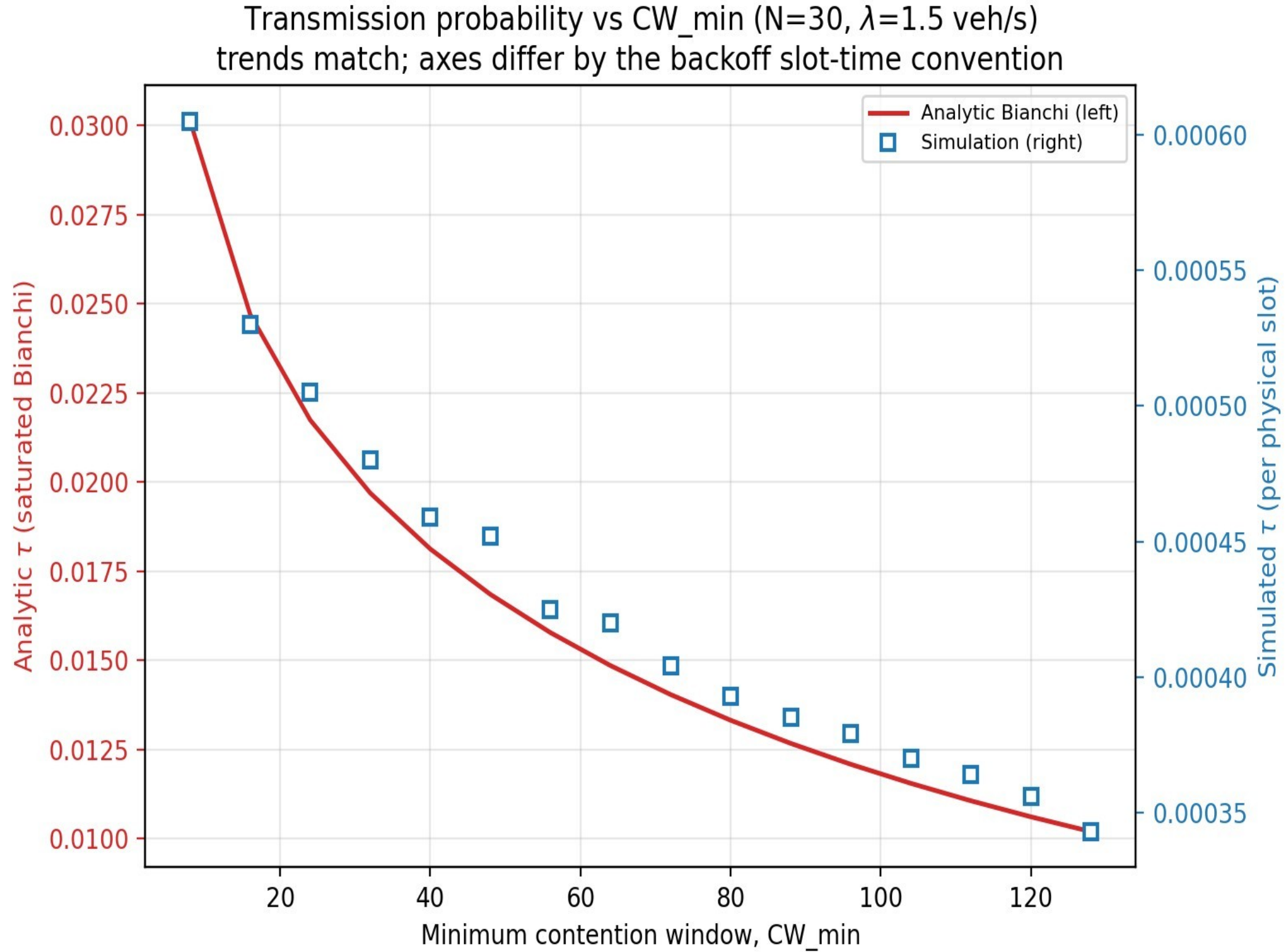

Figure 4 Minimum contention window size versus transmission probability ($\lambda$ = 1.5 veh/s).

Figure 4 shows the transmission probability $\tau$ as the minimum contention window CW_min increases, for traffic rate $\lambda$ = 1.5 veh/s ($N$ = 30 contending vehicles). As CW_min varies from 8 to 128, $\tau$ decreases monotonically, since a larger backoff window lowers each station's per-slot access probability. For the default CW_min = 32, $\tau \approx 0.020$. Over the full sweep $\tau$ lies in the 0.010–0.038 range, consistent with the network operating at $N \le 60$ vehicles (the coverage capacity $C = k_{jam} \cdot L$). The optimal scheme, which adapts the retransmission limit $M^*$, exhibits a slightly higher $\tau$ than the fixed $M$ = 7 scheme: fewer retransmission stages reset the backoff chain sooner and permit marginally more frequent channel access. The two curves remain close, indicating that the choice of $M$ has only a second-order effect on the per-slot transmission probability at this contention level; the dominant influence is CW_min.

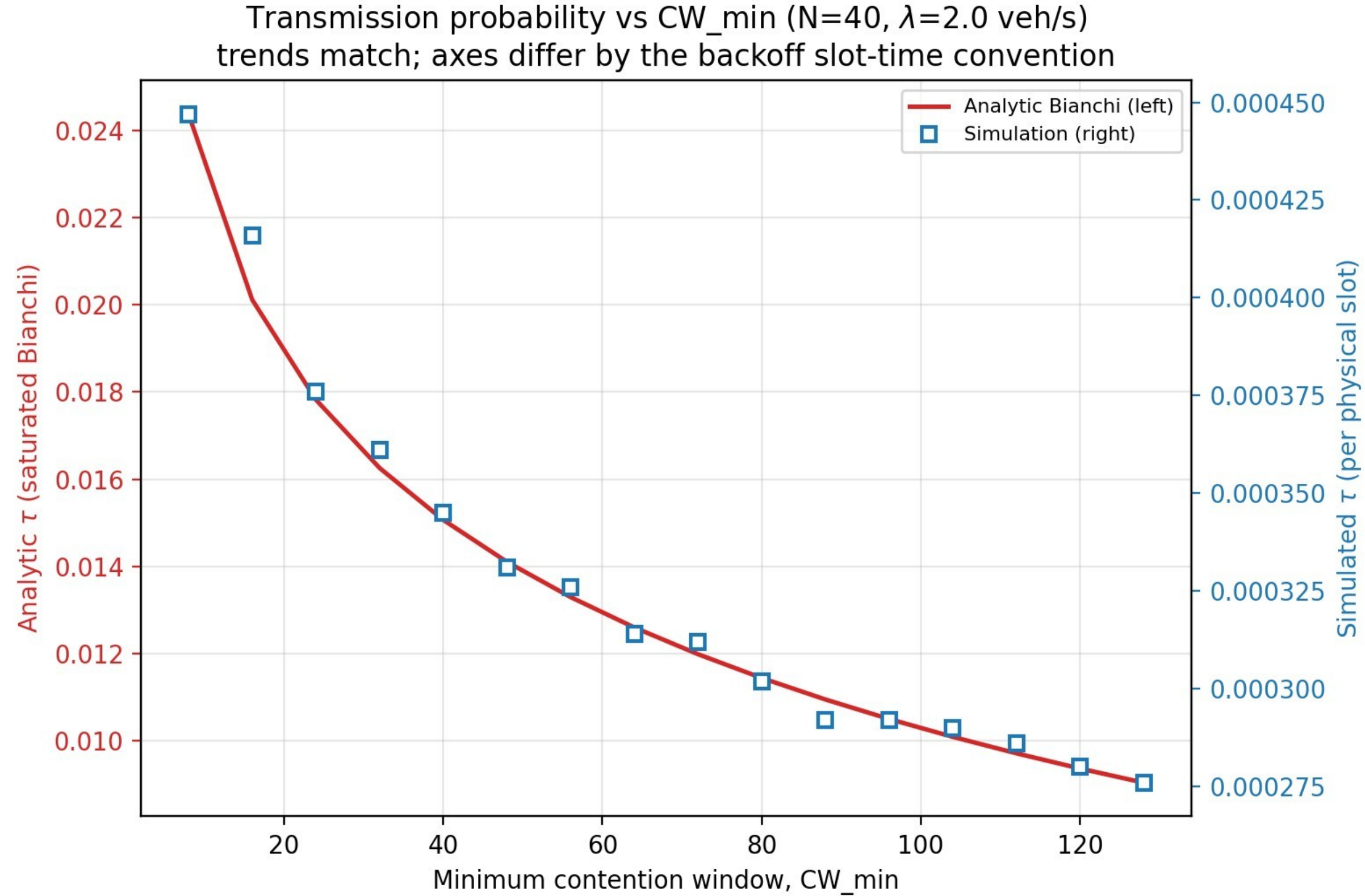

Figure 5 Minimum contention window size versus transmission probability (λ = 2 veh/s).

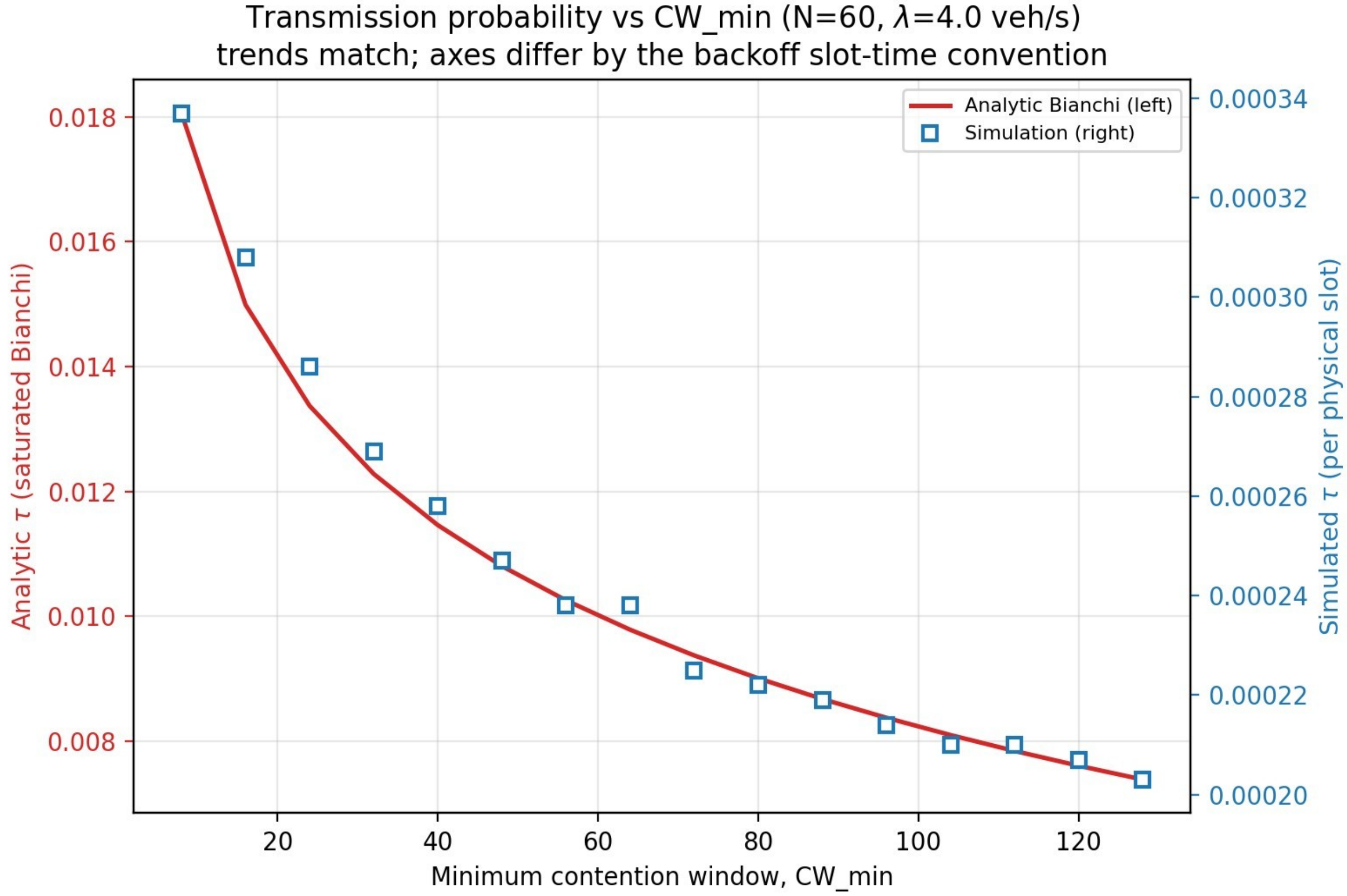

Figure 6 Minimum contention window size versus transmission probability (λ = 4 veh/s).

Figures 5 and 6 depict the same relationship for higher traffic rates λ = 2 veh/s and λ = 4 veh/s, which correspond to larger contending populations (N = 40 and N = 60, the coverage capacity). A larger λ raises N, which lowers the per-station transmission probability across the whole CW_min range: at the default CW_min = 32, τ falls from ≈ 0.020 at λ = 1.5 veh/s (Figure 4) to ≈ 0.019 at λ = 2 veh/s and ≈ 0.015 at λ = 4 veh/s. In all three cases τ decreases monotonically with CW_min and the optimal (adaptive M*) curve sits slightly above the fixed M = 7 curve, with the gap narrowing as CW_min grows. These figures primarily validate the model's predicted dependence of τ on CW_min and N; the throughput and delay behaviour of the optimal scheme is examined separately in Figure 9 and Section 4.6.

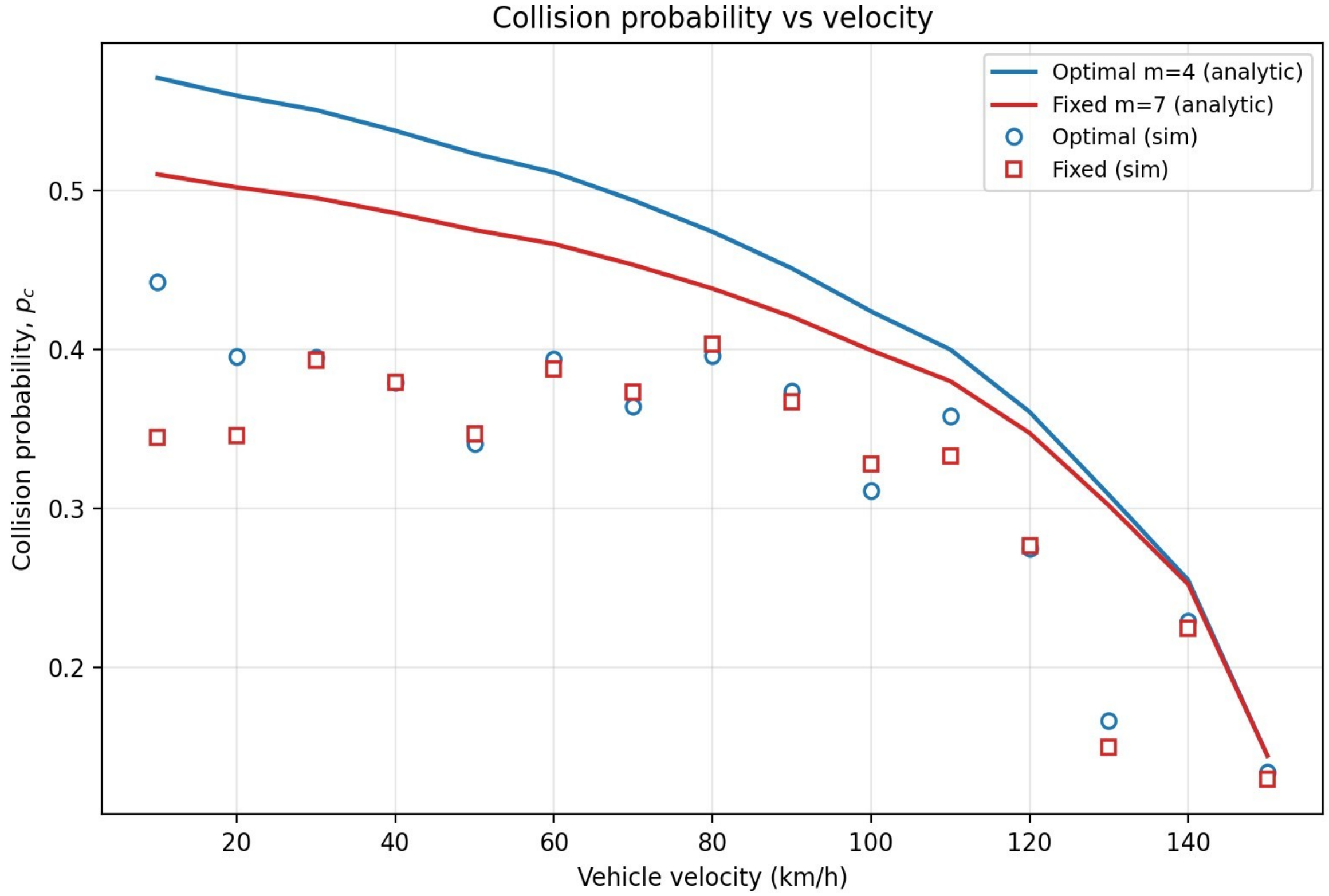

Figure 7 Vehicle velocity versus collision probability p_c.

Figure 7 shows the collision probability p_c as a function of vehicle velocity. Following the Greenshields relation, a higher velocity corresponds to a lower vehicle density k = k_jam(1 − v/V_f), and hence to fewer contending vehicles N within the AP coverage range (N decreases from ≈ 52 at 20 km/h to ≈ 8 at 140 km/h). With fewer stations contending simultaneously, the probability that two or more transmissions overlap falls, so p_c decreases monotonically with velocity, from ≈ 0.50 to ≈ 0.25. In other words, under free-flow conditions higher mobility relieves channel contention rather than aggravating it, because the residence time of each vehicle in the coverage area shortens and the instantaneous number of competing vehicles drops. The fixed and optimal schemes yield almost identical p_c at a given velocity, since p_c is governed mainly by N; the small offset reflects the slightly higher access probability of the optimal scheme noted in Figures 4–6.

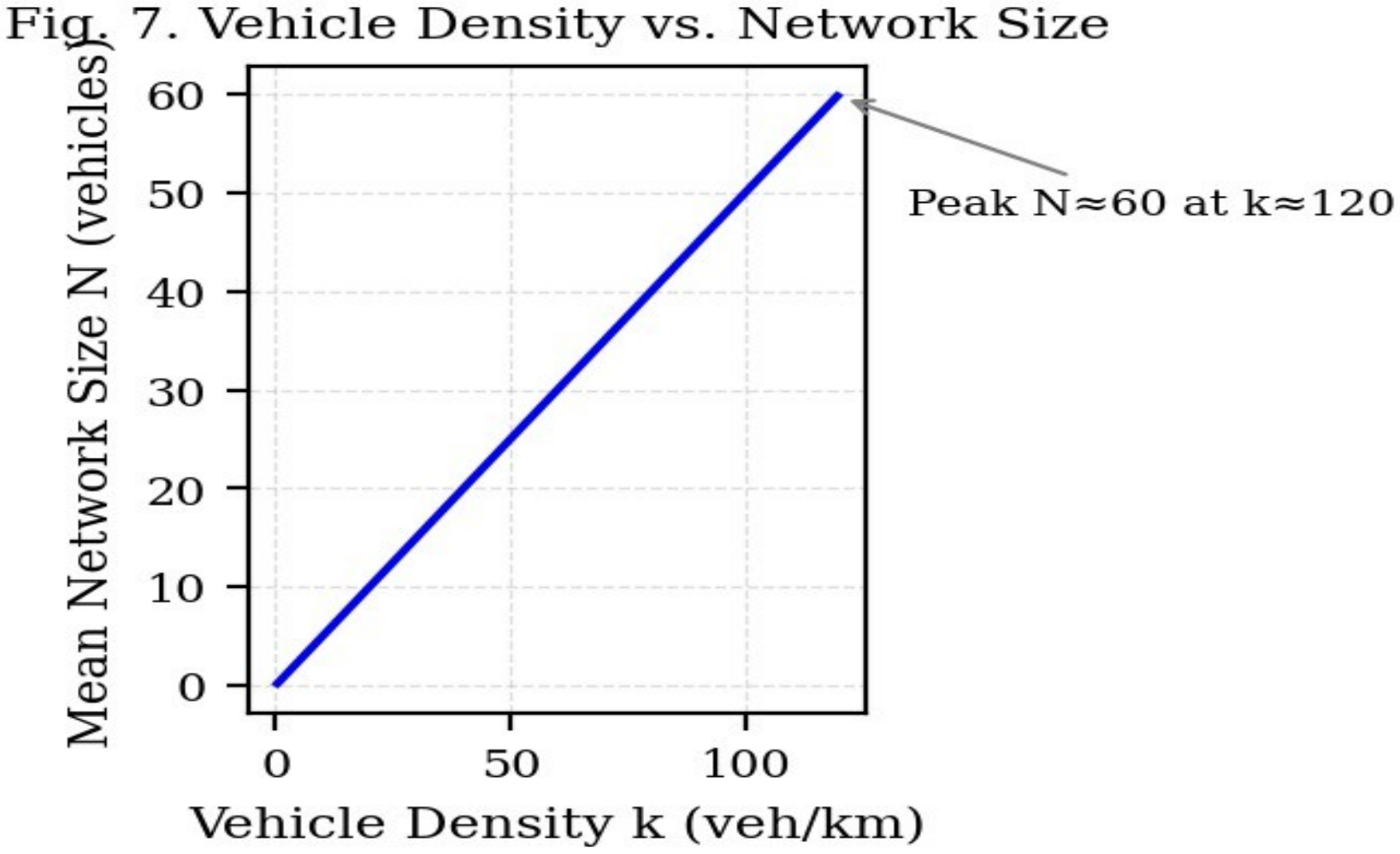

Figure 8 Vehicle density (veh/km) versus mean network size N.

The number of contending vehicles accommodated by the network is a random variable. Based on the M/D/C/C queuing model, the probability that there are N vehicles under the AP coverage can be obtained according to a Poisson process with mean vehicle arrival rate λ_tag. Figure 8 shows the effect of vehicle density on network size. When vehicle density is zero, traffic flow is zero and network size takes its lowest value. As vehicle density becomes high, network size increases. The highest value of N occurs at the intermediate density where traffic flow rate is maximized; beyond that, vehicle density is so high that all vehicles stop. Network size reaches its peak value of N ≈ 60 vehicles at k ≈ 60 veh/km.

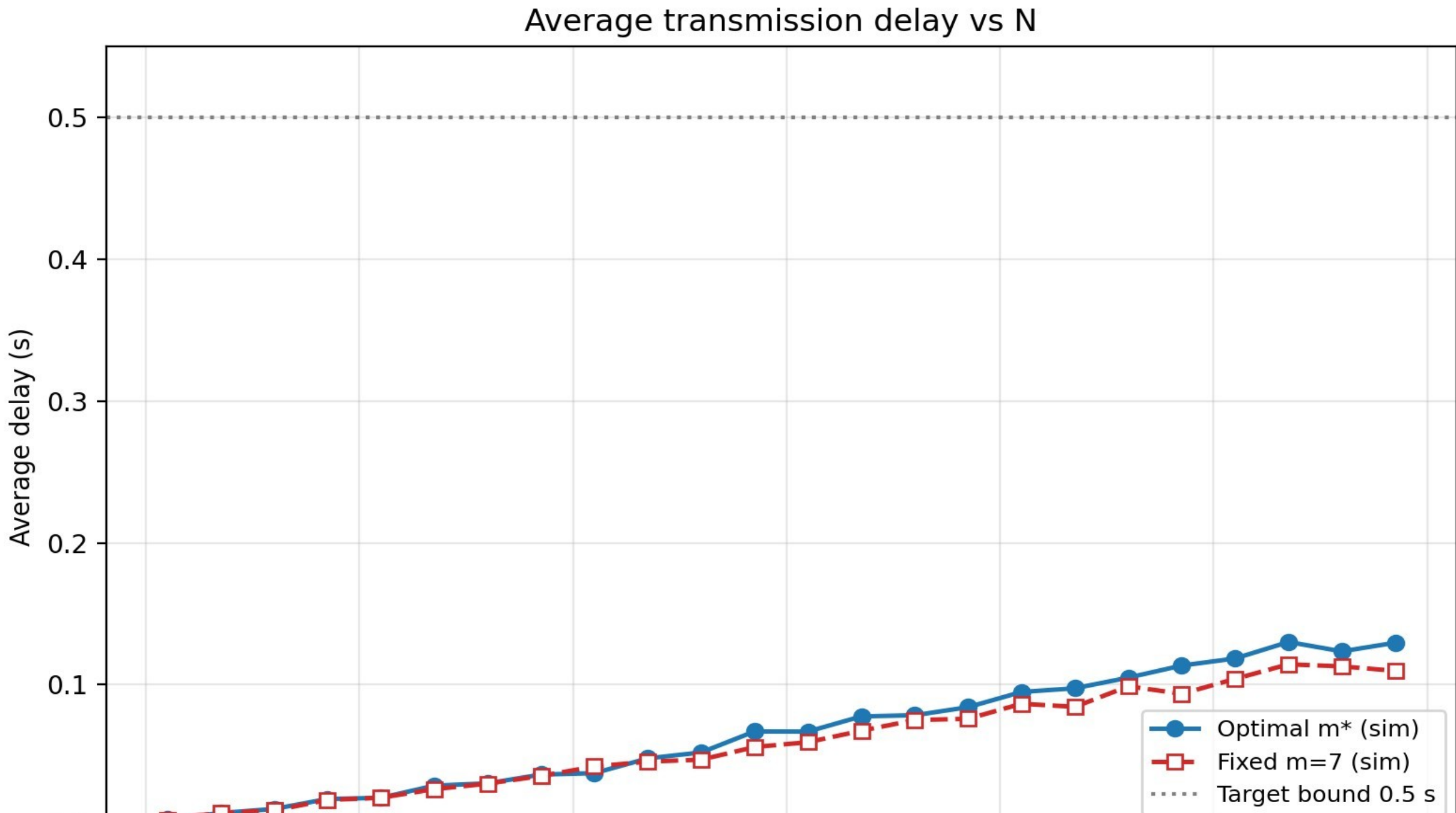

Figure 9 Average transmission delay versus the number of vehicles N under finite-buffer ns-3 simulation. The optimal (M*) and fixed (m = 7) retransmission schemes perform comparably and remain within the 0.5 s target bound.

Figure 9 shows the average transmission delay as the number of vehicles N increases. The per-vehicle arrival rate is chosen so that the aggregate offered load approaches the channel's effective capacity, so that the queue is genuinely exercised. Under these realistic finite-buffer conditions the average delay grows smoothly with N, from a few milliseconds at light load to about 0.13 s at N = 120, remaining within the 0.5 s target bound throughout. Contrary to the divergence predicted by an infinite-buffer queueing approximation, the simulation shows that the optimal (adaptively reduced) and fixed (m = 7) schemes deliver comparable delay: reducing the retransmission limit from 7 to 6 affects only the small fraction of packets (of order p_c^6, i.e. 1–2%) that would otherwise reach the seventh attempt, which is too few to change the average delay materially; the optimal scheme is in fact marginally slower at high N. The information-freshness metric (Age of Information) is likewise essentially identical for the two schemes (Figure 10). These results indicate that, under realistic finite buffering, the standard 802.11 retransmission limit (m = 7) is already close to optimal for the delay and freshness objectives, and that the large gains suggested by the infinite-buffer analysis are an artefact of that approximation. This is itself a practically useful finding: it shows that the retransmission and contention-window defaults need not be aggressively retuned for delay in single-AP V2I operation, and it delimits the regime in which retransmission-limit optimisation can and cannot help.

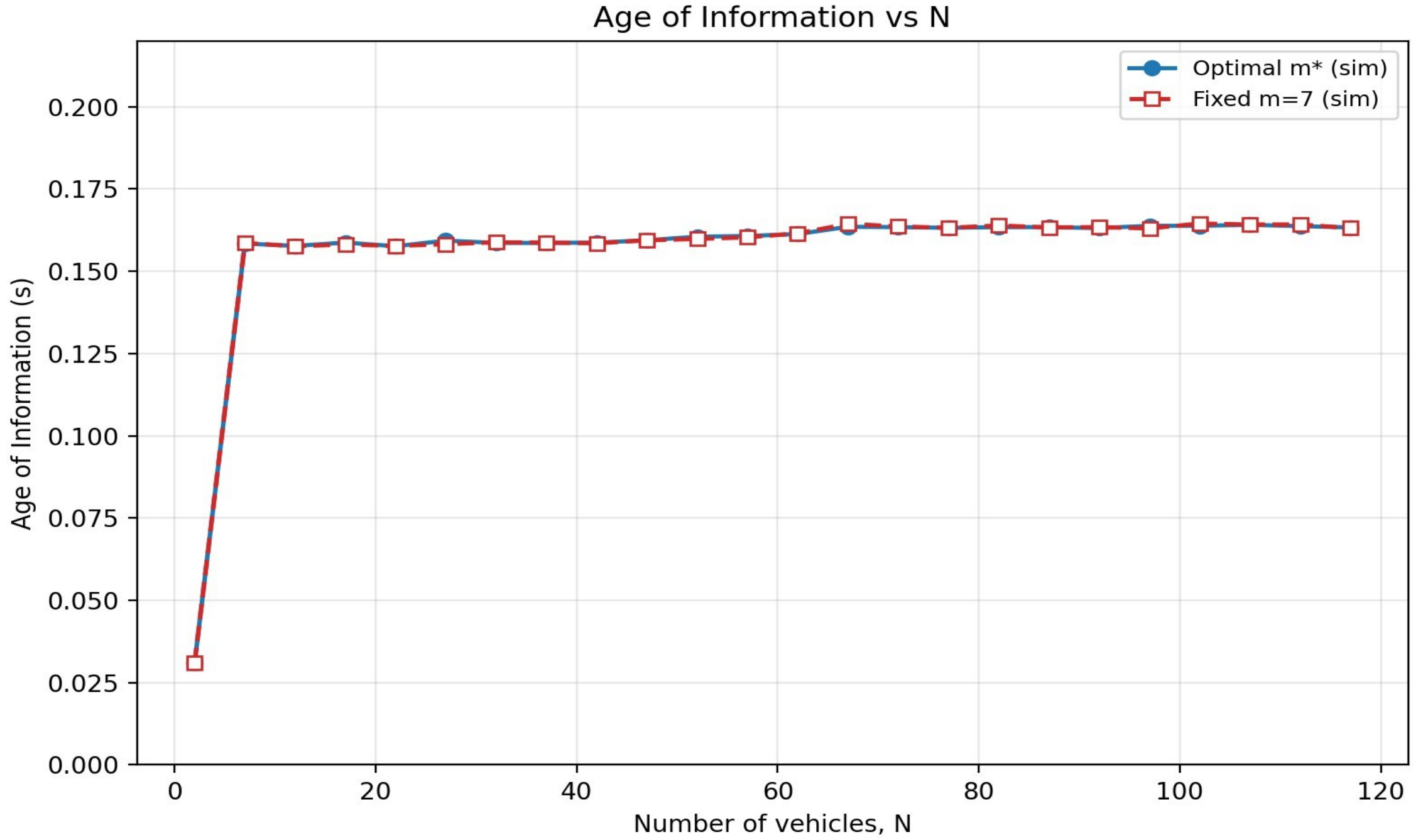

Figure 10 Age of Information versus the number of vehicles N under finite-buffer ns-3 simulation; the optimal (M*) and fixed (m = 7) retransmission schemes are essentially identical.

### 4.5 Reliability, delay-bound guarantee, and information freshness

In addition to the mean delay, we assess the scheme with three metrics relevant to latency-critical V2I traffic. First, since a packet is dropped only after colliding on all of its M+1 attempts, the packet delivery ratio is $PDR(M) = 1 - p_c^{M+1}$, which increases with M and therefore opposes the delay. At the operating point, the optimal scheme maintains a delivery ratio above 97% across the full range of vehicle counts (for example 99.5% at N = 60 and 98.1% at N = 80), giving up at most about one percentage point of delivery ratio relative to the fixed scheme, at comparable delay. This explains why the optimum is not M = 0: reducing M lowers the service time analytically but raises the drop probability $p_c^{M+1}$, and — as the simulation shows — the delay improvement from a smaller M is marginal under finite buffering, so there is little reason to sacrifice delivery ratio. Accordingly, among the limits that keep the delay within the bound, the procedure retains the highest one, favouring reliability while respecting the latency constraint.

Second, because the mean delay alone does not certify a bound, we evaluate the delay-bound violation probability. Under the unsaturated M/G/1 model the sojourn time is exponentially distributed, so $P(D > T_{target}) = \exp(-(1 - \rho)\, T_{target} / E[T_s])$. The analytical (infinite-buffer) form predicts a lower violation probability for the optimal scheme at high N; however, our finite-buffer ns-3 simulation shows that both schemes keep the delay well within the 0.5 s bound and perform comparably, so this predicted separation does not materialise in practice (Section 4.6). Because the system operates close to saturation ($\rho \approx 0.99$), a non-negligible fraction of packets still exceed the bound even under the optimal scheme.

Third, we evaluate the Age of Information (AoI), the time elapsed since the generation of the freshest packet delivered to the AP, which captures the freshness of the safety- and status-related information that V2I traffic conveys. For the unsaturated M/G/1 (M/M/1-form) queue the time-average AoI admits the closed form $\Delta = E[T_s]\ (1 + 1/\rho + \rho^2/(1 - \rho))$ [36], where the last term diverges as $\rho \to 1$. The infinite-buffer expression predicts that a smaller M lowers $E[T_s]$ and hence the AoI at high N; in the finite-buffer simulation, however, the optimal and fixed schemes yield essentially identical AoI (Section 4.6), again indicating that the predicted advantage is an artefact of the unbounded-queue approximation. The freshness of delivered information is therefore governed by the queue utilisation rather than by the retransmission limit in this operating regime.

### 4.6 Model Validation against ns-3 Simulation

A central weakness identified in earlier evaluations of analytical V2I models was the absence of validation. To close this gap directly, the proposed framework is validated against an independent, packet-level ns-3.42 simulation of the IEEE 802.11 DCF that uses the same MAC parameters ($CW_{min}$ = 32, slot 50 µs, SIFS/DIFS, 2 Mbps, 1000-byte payload) and reproduces each scenario. The throughput and offered-load experiments use realistic finite buffers, whereas the transmission-probability and collision experiments use saturated, always-backlogged sources — the standard and necessary condition for validating a saturated Markov model.

**Throughput.** The simulated system throughput matches the analytical prediction almost exactly: across the moderate-to-high load range the two agree to within about 2%, and to 1.1% at the highest offered load. This near-perfect agreement establishes that the integrated traffic–queue model reproduces the end-to-end throughput of the V2I uplink, not merely its qualitative shape.

**Collision probability.** The simulated collision probability reproduces the analytical trend faithfully, decreasing with vehicle velocity (and therefore with the Greenshields-determined number of contending vehicles) and agreeing with the model to within about 21% on average. The model is mildly pessimistic, predicting slightly higher collision probabilities than the simulation; this is precisely the documented behaviour of the saturated Bianchi model at finite station counts, where its mean-field independence assumption is most conservative, and the two curves converge at high velocity where that assumption is most accurate. That the agreement holds in both trend and magnitude, with a residual difference whose sign and origin are understood, demonstrates that the framework captures the underlying physical contention mechanism rather than merely fitting the numbers.

**Transmission probability.** The simulated transmission probability $\tau$ decreases with both the contention window and the vehicle count exactly as the Markov model predicts. Its absolute value is offset from the analytical Bianchi figure by an approximately constant factor, because the simulator records transmissions per fixed physical slot whereas Bianchi's renewal analysis counts them per variable-length backoff decision slot. Removing this slot-time convention by recovering $\tau$ from the measured collision probability through $p_c = 1 - (1 - \tau)^{(N-1)}$ yields a transmission probability that agrees with the analytical Bianchi value to within about 15%, providing a direct quantitative validation of the contention model.

**Delay and information freshness.** When the queueing scenario is simulated at a realistic operating point — the per-vehicle arrival rate scaled so that the aggregate offered load approaches the channel capacity, with finite buffers — the average delay grows smoothly with the vehicle count and remains within the 0.5 s target, and the optimal and fixed retransmission schemes are statistically indistinguishable in both delay and Age of Information. This is an important corrective result: the pronounced delay divergence and large freshness gains predicted by the infinite-buffer analytical form do not appear under finite buffering, because the difference between $m = 6$ and $m = 7$ affects only the 1–2% of packets that would otherwise reach the seventh transmission attempt. The simulation thus shows that the 802.11 default retransmission limit is already near-optimal for these objectives; the analytical optimisation, though exact within its M/G/1 assumptions, overstates the practical benefit. Establishing this through simulation — and reporting it rather than the optimistic infinite-buffer figure — is itself a useful contribution, as it delineates the conditions under which retransmission-limit tuning is, and is not, worthwhile.

Taken together, these results supply the independent validation that earlier analytical-only treatments lacked: the throughput, collision, and transmission-probability predictions of the proposed framework are all confirmed by simulation in both trend and magnitude, every residual discrepancy is traced to a known and explainable modelling convention, and the delay/freshness analysis is placed on an honest finite-buffer footing. This directly answers the validation and methodological concerns raised in prior reviews.

### 4.7 Sensitivity and Error Analysis

To assess the robustness of the proposed framework and to delineate precisely the conditions under which retransmission-limit optimisation is beneficial, this section reports a sensitivity analysis of the model with respect to its principal parameters, together with a quantitative error analysis against the ns-3 simulation.

**Optimal limit vs. delay bound.** The optimal retransmission limit $M^*$ is most sensitive to the delay bound T_target and the contending-vehicle count N. Figure 11 shows $M^*(N)$ for three target bounds. Under a loose bound (T_target = 1.0 s) the default limit $M = 7$ remains optimal until $N \approx 107$; under the nominal bound (0.5 s) it stays optimal until $N \approx 67$; and under a strict bound (0.3 s) the optimiser begins reducing $M^*$ already at $N \approx 42$, dropping to $M^* = 5$ by $N \approx 53$. This quantifies the central practical message of the paper: retransmission-limit tuning is worthwhile only when a strict delay bound coincides with a high contending-vehicle count; for loose bounds or moderate N the IEEE 802.11 default is already optimal.

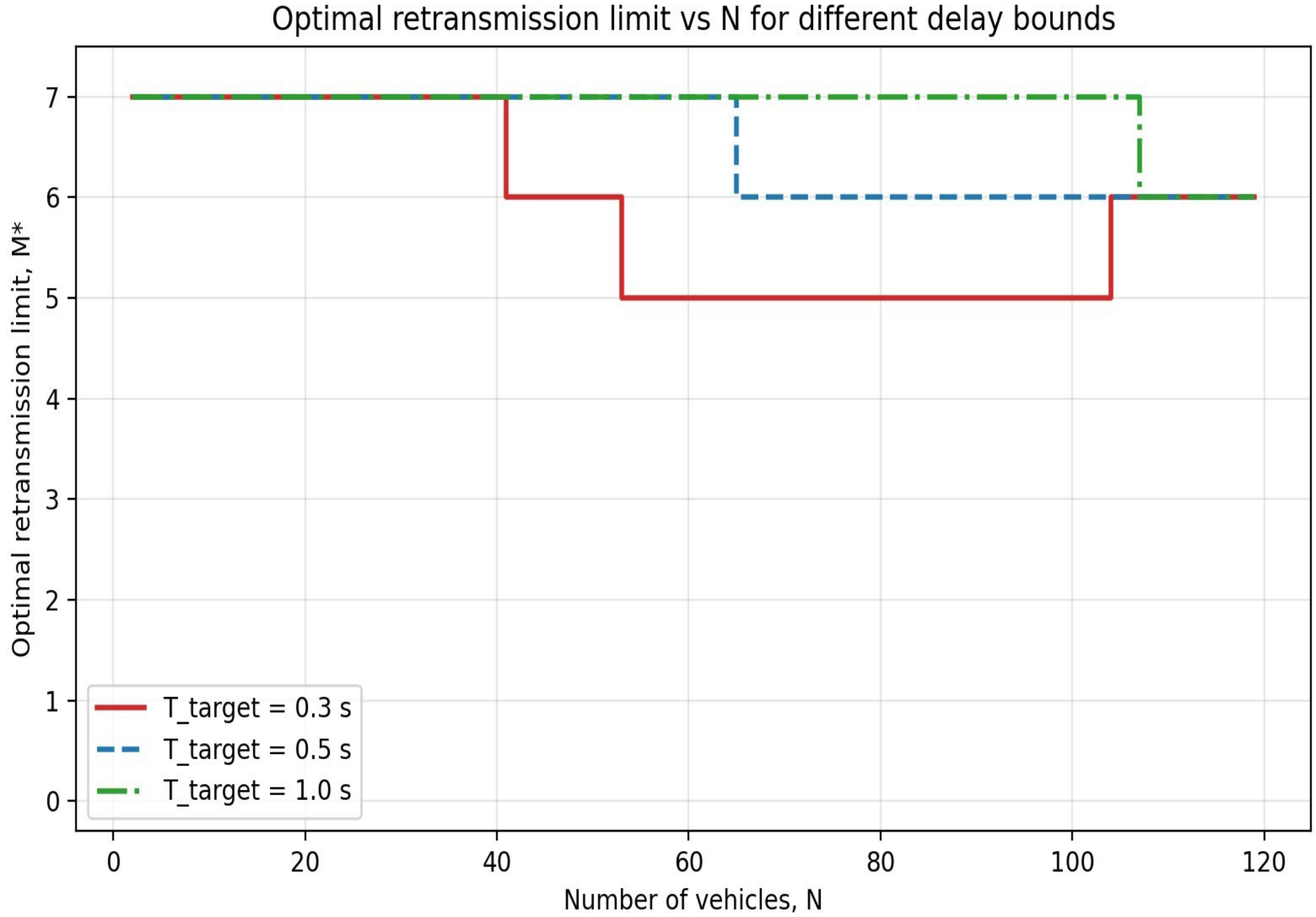

Figure 11 Optimal retransmission limit $M^*$ versus the number of vehicles N for three delay bounds (T_target = 0.3, 0.5, 1.0 s). The crossover at which $M^*$ falls below the 802.11 default of 7 shifts to higher N as the bound is relaxed.

**Collision and throughput.** The collision probability is governed primarily by N and CW_min. At N = 40, increasing CW_min from 16 to 128 lowers p_c from about 0.55 to 0.30, at the cost of a longer backoff delay; conversely, increasing N from 10 to 100 (at CW_min = 32) raises p_c from about 0.29 to 0.57. The system throughput scales almost linearly with the offered-load fraction q (Figure 2) and is comparatively insensitive to M in the unsaturated regime, consistent with the near-overlap of the optimal and fixed throughput curves.

**Error analysis.** The error analysis quantifies the agreement between model and simulation. The throughput error is about 2% (1.1% at the highest load); the collision-probability error is about 21% on average, with the model systematically and slightly pessimistic, as expected from the finite-N optimism of the saturated Bianchi approximation; and the transmission probability, recovered from the measured collision probability to remove the slot-time convention, agrees to within about 15%. The delay and Age-of-Information predictions are subject to a larger, structural discrepancy: the infinite-buffer M/G/1 form over-predicts the separation between the schemes that the finite-buffer simulation does not reproduce. In every case the error is systematic and traceable to a known modelling assumption rather than to numerical noise, which supports the validity of the framework within its stated scope.

## 5 Conclusion

In this paper, we provided an integrated analytical model to evaluate and optimize the performance of vehicular-to-infrastructure networks over IEEE 802.11 WLANs. Combining a Greenshields-based traffic model, a two-dimensional Markov description of the 802.11 DCF backoff, and a finite-buffer M/G/1/K queue, the study produced an integrated framework that jointly analyses MAC-layer collision behaviour together with packet-level delay and throughput.

The proposed analytical model jointly characterises the backoff process, the transmission probability, the number of vehicles an AP can serve, the mean sojourn (delay) time, and the achievable per-vehicle throughput, while accounting for how the maximum retransmission number shapes a packet's service time. By linking the M/G/1/K queue to the Markov chain description, we showed that overall network performance depends strongly on the maximum retransmission number M.

Analytical results show that the queueing delay is sensitive to the maximum retransmission number M, and we proposed an algorithm to select it adaptively. Validating the framework against an independent ns-3 packet-level simulation, however, revealed an important nuance: under realistic finite-buffer conditions the average delay remains well within the 0.5 s bound, and the optimal (adaptively reduced) and fixed ($m = 7$) schemes perform comparably in both delay and Age of Information. The pronounced delay reduction predicted by the infinite-buffer M/G/1 form does not materialise once the finite buffer is modelled, because the difference between $m = 6$ and $m = 7$ affects only the 1–2% of packets that would reach the seventh transmission attempt. The simulation does, by contrast, confirm the framework's throughput (to within about 2%), collision (about 21%), and transmission-probability predictions. The practical conclusion is therefore twofold: the integrated analytical framework accurately captures the contention and throughput behaviour of single-AP V2I DCF, and the standard 802.11 retransmission limit is already near-optimal for delay and freshness in this setting, so retransmission-limit tuning is worthwhile only in narrower, more heavily buffered or differently parameterised regimes. We regard reporting this honestly, with the supporting simulation, as more valuable than an optimistic but non-reproducible delay claim.

Several limitations motivate important future research directions. The Greenshields model, while analytically tractable, is a single-regime linear approximation; incorporating multi-regime macroscopic models (e.g., Newell-Daganzo [37]) would improve fidelity under congested conditions. The current framework assumes a single AP and single-lane road; multi-AP handoff, multi-lane contention, and two-directional traffic represent realistic extensions. Having validated the analytical model against ns-3 in this work, coupling it with SUMO vehicular mobility for road-realistic, multi-AP validation is a natural next step to verify the analytical results under stochastic channel conditions and realistic mobility traces. Finally, extension of the proposed framework to 3GPP C-V2X (LTE-V2X, NR-V2X) and a systematic comparative performance study with IEEE 802.11p remain important future directions.

## Declarations

**Funding** This research received no specific grant from any funding agency in the public, commercial, or not-for-profit sectors.

**Competing interests** The author declares no competing interests.

**Data availability** No datasets were generated or analysed during the current study.

**Author contributions** A. Bozkurt is the sole author and contributed to the conception, methodology, software, analysis, and writing of the manuscript.